\documentstyle[aps,twocolumn,epsfig,prl]{revtex}
\begin{document}
\draft
\flushbottom
\twocolumn[
\hsize\textwidth\columnwidth\hsize\csname @twocolumnfalse\endcsname

\title{A Lattice-Spin Mechanism in Colossal Magnetoresistant Manganites}
\author{
J. A. Verg\'es$^{(1)}$, 
V. Mart\'{\i}n-Mayor$^{(2)}$ and 
L. Brey$^{(1)}$}
\address{
$^1$Instituto de Ciencia de Materiales (CSIC). Cantoblanco,
28049 Madrid. Spain.\\
$^2$ Dipartimento di Fisica, Universit\`a di Roma ``La
Sapienza'', Piazzale Aldo Moro 2, 00185 Roma, Italy \\ INFN sezione di
Roma.
} 
\date{\today}
\maketitle 

\begin{abstract}
We present a single-orbital double-exchange model, coupled with {\bf
cooperative} phonons (the so called breathing-modes of the oxygen
octahedra in manganites).  The model is studied with Monte Carlo
simulations. For a finite range of doping and coupling constants, a
first-order Metal-Insulator phase transition is found, that coincides
with the Paramagnetic-Ferromagnetic phase transition. The insulating
state is due to the self-trapping of every carrier within an oxygen
octahedron distortion.
\end{abstract}

]
\narrowtext
\tightenlines

Mixed-valence manganites have attracted much attention lately, since
they undergo a transition from a ferromagnetic (F) to a paramagnetic (P)
state accompanied by a metal (M) to insulator (I) transition.
The Double Exchange (DE) mechanism\cite{de} plays a major role to explain
the magnetic transition, but the mechanism responsible of the M-I
transition is not fully understood\cite{rvw}. In the DE model, due to
a very strong Hund's coupling, the carriers
are strongly ferromagnetically coupled to the Mn core spins producing
a modulation of the hopping amplitude between Mn ions.  Recently a
growing number of experiments\cite{jaime99,heffner00,adams00,wu01}
support the idea that phase separation is important in
manganites\cite{arovas99,dagottorev,burgy}.  In this
state, a phase is metallic and the other insulating.  The theoretical
challenge is thus to find a FM---PI phase transition of the first
order.  Whereas it is widely assumed that the metallic phase is the
ferromagnetic DE phase, it is not clear what is the origin of the gap
in the insulating phase.  For doping level near half filling and low critical
temperature ($T$) materials, it has been proposed and widely accepted
that the insulator is a charge-orbital ordered phase with a strong short
range order.  At lower doping levels the insulating phase is supposed to be
some kind of polaron gas\cite{millis1,roder} (polaron meaning a
lattice-spin object) or polaron lattice phase\cite{mathur,mizokawa}.

An attractive picture for polaron formation was presented in
Refs.~\cite{millis1,roder}: it was observed that the importance of the
electron-lattice coupling is given by its ratio with the carriers
kinetic energy. Since within the DE mechanism the kinetic energy
decreases upon heating, it was proposed that localized polarons are formed
at the F-P transition giving rise to a M-I transition.  Millis
{\em et al.} \cite{millis1} used the dynamical mean field method to
study the coupling of the carriers to local Jahn-Teller distortions
and to the Mn core spins. At half filling an I-M transition was found
close to the Curie $T$. It was shown that the electron-phonon coupling
could be tuned to reproduce the $T$-dependence of the resistivity of
several manganites. Unfortunately, this approach presented several
caveats: the M-I transition was only found at half-filling, phonons
are treated classically, and (most important) intersite phonon
correlations were not considered.

In this Letter, we consider a single-orbital (s-wave) DE model coupled
with phonons.  The model will be kept as simple as possible, since our
scope is to shed some light on the mechanism behind the coupling
between the M-I transition and the Curie temperature.  Both core-spins
and phonons are treated as classical variables (for spins this is a
controlled approximation~\cite{classical}).  The lattice distortion we
study is the deformation of the oxygen octahedra around Mn sites.
The coupling of these modes with charge carriers is expected to be at
least as large as the one producing Jahn-Teller distortions\cite{millis2}.
The contraction ({\em breathing}) of a MnO$_6$ octahedron, implies a
volume growth of its neighbors (but do not change the total lattice
volume, as would be needed to study magnetostriction
effects~\cite{deTeresa}). Thus this mode is strongly cooperative, and
not suitable for Mean-Field studies. A Monte Carlo (MC) investigation
is therefore performed.
The super-exchange antiferromagnetic coupling between the core
spins~\cite{Alonsos} is neglected in our model.
Also, as we mentioned above, we work with a
single s-orbital per site and we do not consider the two degenerated
${\bf e_g}$ orbitals which are crucial to understand the magnetic
phase-diagram beyond half-filling~\cite{Brink99}. Thus, our model is
to be regarded as a model
for materials like La$_{1-x}$Ca$_x$MnO$_3$ in the $0.15<x<0.4$ regime
where the magnetoresistance is largest, and the only experimentally
relevant magnetic phases are F and P~\cite{rvw}.  In spite of its
simplicity, the model presents a first-order M-I phase transition,
that coincides with the P-F phase transition {\em for a finite-range
of doping} and coupling constants, in contrast with previous
work~\cite{millis1}.

The model Hamiltonian contains the Mn ${\bf e_g}$ itinerant carriers
coupled to the ${\bf t_g}$ Mn core spins, and phonons:
\begin{equation}
H = H_{\mathrm{KE}}+ H _{\mathrm{Hund}} + H _{\mathrm{el-ph}}
+ H_{\mathrm ph} \,  .
\label{hamil}
\end{equation}
Here $H_{\mathrm{KE}}$ is the kinetic energy of the carriers hopping
between Mn atoms that form a simple-cubic lattice, $H_{\mathrm{Hund}}$
is the Hund interaction, $H _{\mathrm{el-ph}}$ is the lattice-carrier
coupling energy, while $H_{\mathrm{ph}}$ represents the crystal
elastic energy. The Hund interaction is very large in
manganese oxides and at each place the carrier spin is forced to be
parallel to the core spin, which allows to reduce 
$H_{\mathrm{KE}}+H_{\mathrm{Hund}}$ to the DE
Hamiltonian~\cite{dagottorev}:
\begin{equation}
H_{\mathrm{DE}} =\,\sum_{\langle i,j\rangle}\  \left(
 {\cal T}( \mbox{\boldmath$S$}_i,\mbox{\boldmath$S$}_j )\ 
c_i^\dag c_j \ +\ {\mathrm{h.c.}}\right)\,.
\label{hamilde}
\end{equation}
Here $c ^+ _i$ creates an electron at place $i$ with spin parallel to
the core spin at $i$, $t$ is the hopping amplitude between first
neighbors ions and ${\cal T}(
\mbox{\boldmath$S$}_i,\mbox{\boldmath$S$}_j )=
-t\,[\cos\frac{\theta_i}{2}\cos\frac{\theta_j}{2}+
\sin\frac{\theta_i}{2}\sin\frac{\theta_j}{2} {\mathrm e}^{{\mathrm
i}(\varphi_i-\varphi_j)}]$, $\theta _i$ and $\phi _i$ being the polar
coordinates of the core spin at site $i$, $\mbox{\boldmath$S$}_i$.
For the electron-phonon coupling we consider the distortions of the
MnO$_6$ octahedron formed by the six oxygens surrounding the Mn ions.
The oxygens are located at the center of the edges of the cubic
lattice formed by Mn atoms. Each oxygen is allowed to move along the
edge on which it is located. The distortions of the six oxygens
surrounding a Mn at site $i$ are given by $u_{ i,\pm \alpha}$ where
$\alpha$ run over $x$, $y$, $z$.  The size fluctuations of the MnO$_6$
octahedra are coupled to charge fluctuations in the Mn through the
electron-phonon interaction,
\begin{equation}
H_{\mathrm{e-ph}} = - \lambda t\sum _{i, \alpha}
\left ( u _{i,- \alpha} - u _{i,\alpha} \right ) c ^+ _ i c _i \, \, \, ,
\label{elph}
\end{equation}
where $\lambda$ is the electron-phonon coupling.  This interaction
tends to produce lattice distortions. This tendency is opposed by the
stiffness of the Mn-O bonds:
\begin{equation}
H_{\mathrm{ph}}= t \sum _{i,\alpha} \left ( u _{i,\alpha} \right  ) ^  2 \, \, \, .
\label{phonon}
\end{equation}
\begin{figure}
\epsfig{file=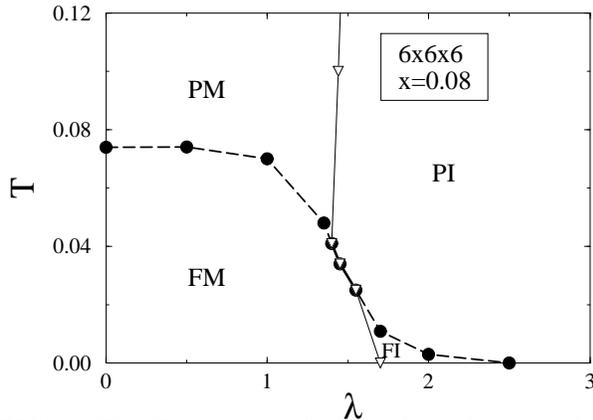, width=\columnwidth, angle=0}
\caption{The $T$ (in units of $t$)---$\lambda$ phase diagram of the model (\ref{hamil}), obtained from MC simulations
on a $6^3$ lattice with $x=0.08$ (see text).  Filled dots: P-F
critical $T$.  Open
triangles: M-I critical $T$.}
\label{Fig1}
\end{figure}
One can get some intuition about the physics of our model considering
the limit of very few carriers. If the carrier-lattice coupling is
strong, one can gain enough electronic energy by contracting an oxygen
octahedron (thus localizing a carrier: a {\em polaron}) to compensate
the high price in elastic energy. For the fully spin-polarized
lattice, one finds (using e.g. the techniques of
Ref.~\cite{Vetri-reticoli}) $\lambda^{\mathrm{threshold}}= 1.91$.  The
carrier is localized at the polaron's center with a $96\%$
probability.  If one repeats the calculation for the P phase, using
deGennes' virtual-crystal approximation~\cite{DeGennes}, finds
$\lambda^{\mathrm{threshold}}\approx 1.56$ (the P polaron is also at
its center with $96\%$ probability). Thus for $\lambda$ in between both
thresholds, one expects that at the F-P phase transition, every
carrier will form a strongly localized polaron upon heating. The system
becomes insulating due to the formation of a fully occupied band
separated from upper states by a gap.
This picture is largely confirmed by the MC simulation of the model.
\begin{figure}
\epsfig{file=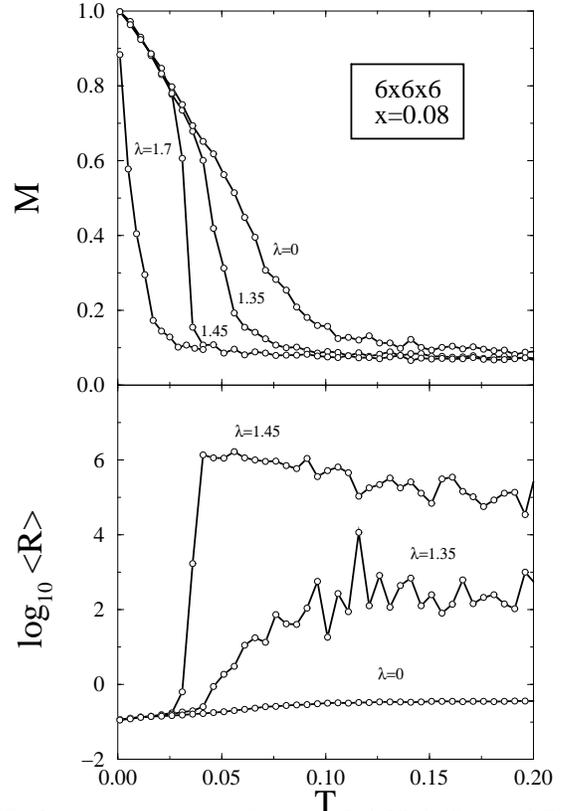, width=\columnwidth, angle=0}
\caption{Top: $M$ vs.
$T$ for $\lambda=0,1.35,1.45$ and $1.7$, for the
MC simulations in Fig.~\ref{Fig1}. Bottom: (decimal) logarithm of 
the average d.c. resistance (a.u.) vs. $T$, from the same
simulations ($\lambda=1.7$ data are
out range).}
\label{Fig2}
\end{figure}
For calculating the $T$-dependent phase diagrams, we perform MC
simulations on the classical variables: the core spins
$\mbox{\boldmath$S$}_i$ and the oxygen displacements ($u_{i,\alpha}$).
The simulations are done in $N \times N \times N$ lattices with
periodic boundary conditions, using an standard Metropolis
algorithm. The kinetic energy of the carriers is calculated by
diagonalizing the electron Hamiltonian at each Metropolis step. The
diagonalization CPU cost grows like $N^6$ and has limited us to
$N=6$~\cite{Nota}.  We have used the $N=4$ results to check for
finite-size effects.
Altghough both absolute values and details sometimes change for $N=4$ simulations,
all issues discussed in this letter remain valid.
The Fermi temperature of the carriers is much
higher than other $T$ in the system and we assume the carriers to be
at zero $T$\cite{calderon}. We calculate the thermal average of
different physical quantities; the absolute value of the
$\mbox{\boldmath$S$}_i$ polarization, $M$, the electronic energy
difference between the lowest energy empty state and the highest
energy occupied one, $E_{\mathrm gap}$, the standard deviation of the (spatial)
probability distribution of the MnO$_6$ octahedra volume, $\Delta
V_{\mathrm rms}$, and the electronic density of states, $\rho (\omega)$.
We also measure
the average d.c. resistance of the system, calculating the
resistance of the $N \times N \times N$ cubic lattice connected to two
semi-infinite perfect leads\cite{verges,calderon2} using the standard
Kubo formula\cite{bastin}.
\begin{figure}
\epsfig{file=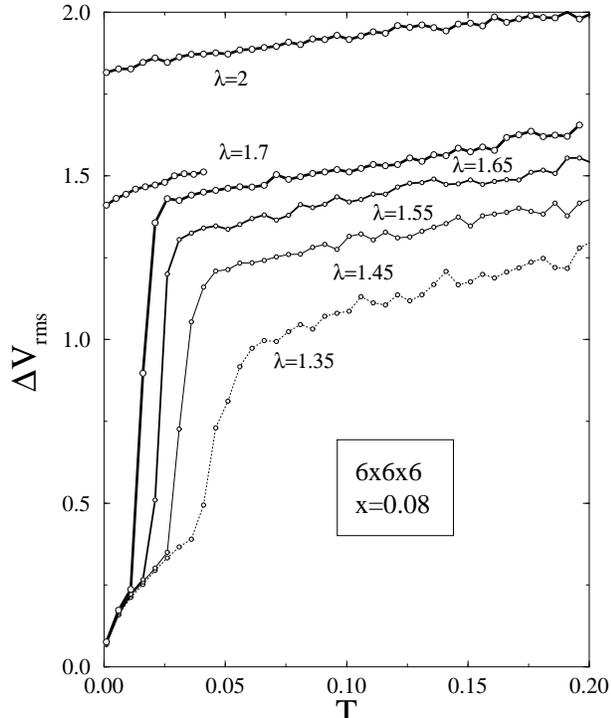, width=\columnwidth, angle=0}
\caption{Standard deviation of the (lattice) distribution of the
volume of the MnO$_6$ octahedra vs. $T$, for the
simulations in Fig.~\ref{Fig1}.
}
\label{Fig3}
\end{figure}
Fig.~\ref{Fig1} shows the phase diagram, $\lambda$ vs. $T$,  for
a $N=$6 cubic lattice with 17 electrons, i.e. $x \simeq$0.08.
We use the criterion that the system is M/I when the
d.c. resistance increases/decreases with $T$.  The phase diagram
contains four phases: FM, FI, PM and PI.  As expected for small
$\lambda$, polarons are not formed by the (small) lattice distortions,
the system being metallic at all $T$~\cite{calderon,calderon2} (see
the growing behavior of the d.c. resistance upon heating in
Fig.~\ref{Fig2}---bottom).  When $T$ grows there is a second order
FM-PM transition, (see the smooth temperature behavior of $M$ in the
top of Fig.~\ref{Fig2}).  At $\lambda$=0, we found the Curie $T$ at
$0.072t$, in agreement with previous MC simulations at $x$=0.08
\cite{calderon}.  Notice that thermodynamic quantities are even
functions of $\lambda$ (because of the symmetry $\lambda\to -\lambda$,
$u_{i,\alpha}\to -u_{i,\alpha}$ in Eq.(\ref{hamil})), and thus for
small couplings they quadratically depend on $\lambda$. This can be
checked for the Curie $T$ in Fig.~\ref{Fig1}.  

\begin{figure}
\epsfig{file=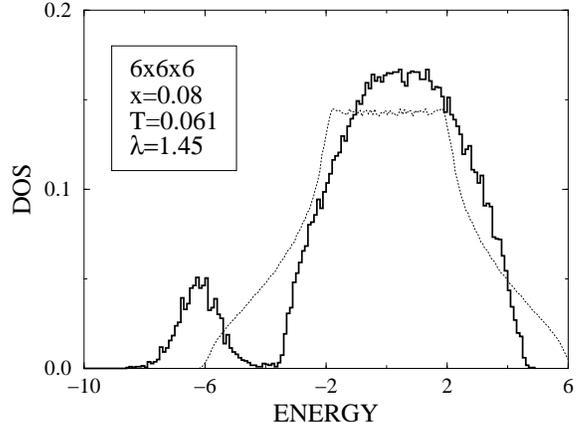, width=\columnwidth, angle=0}
\caption{Thermal average of the electronic density of states for
$\lambda=1.45$, $T=0.061$ (PI phase on Fig.~\ref{Fig1}).  The dashed
line is the density of states of the fully spin-polarized system,
without octahedra distortions.}
\label{Fig4}
\end{figure}

At intermediate $\lambda$, new behavior is expected. In the P phase
(that has higher energy than the F phase), all carriers form polarons
and the system is an insulator,
while in the F phase there are not polarons and the system is metallic.
Upon heating, the degenerated electronic system
undergoes a phase transition at the F-P transition, with a sharp
change in electronic energy. A first-order F-P phase transition with
growing $T$ is obtained, that coincides with a M-I transition.
Indeed, Fig.~\ref{Fig1} shows how the P-M and I-M transition lines
merge for $1.4<\lambda<1.65$. $M$ changes abruptly at the phase
transition (Fig.~\ref{Fig2}, top) while the d.c. resistance grows by a
factor $10^6$ (Fig.~\ref{Fig2}, bottom) and becomes a decreasing
function of $T$. 
First-order P-F transitions are experimentally found in manganites
as the strength of the electron-phonon coupling increases\cite{mira}. 
For slightly smaller values of $\lambda$
($\lambda=1.35 <\lambda_{\mathrm{c}}=1.4$) the d.c. resistance always
grows with $T$, the system being M, and the F-P phase transition is
continuous (Fig.~\ref{Fig2}, top and bottom).  Polaron formation
can also be seen in the spatial distribution of volumes of the MnO$_6$
octahedra that becomes very inhomogeneous: on polarons, octahedra
are small while in most sites the volume is uniform. Note the sharp
change of $\Delta V_{\mathrm rms}$ at the critical line for $1.4<\lambda<1.7$
(Fig.~\ref{Fig3}), and the smoother $T$-dependence for
$\lambda=1.35$. Notice also (Fig.~\ref{Fig4}) the gap in
$\rho(\omega)$ for $1.4<\lambda<1.7$ (the Fermi level is in the gap).
For large $\lambda$, polarons exist in both the P and F phases.  The
F-P and the M-I transitions decouple for $\lambda>1.65$
(Fig.~\ref{Fig1}). The F-P one is again continuous: $M$ evolves
smoothly with $T$ (Fig.~\ref{Fig2}, top), and the system has polarons
at the lowest $T$ (Fig.~\ref{Fig3}), being an I. The F phase in a DE
system with a fully occupied band is somehow unconventional, because the mean
carriers energy is at the band center which is usually spin independent.
This is not the case for our model. The difference between
the positions of the (polaronic) band center of the fully polarized
and unpolarized systems can be calculated as before, with deGennes'
virtual-crystal approximation. The energy difference for large
$\lambda$ is roughly $-1.08 t /\lambda^2$ which explains the (small)
ferromagnetic interaction that decreases with increasing $\lambda$.

\begin{figure}
\epsfig{file=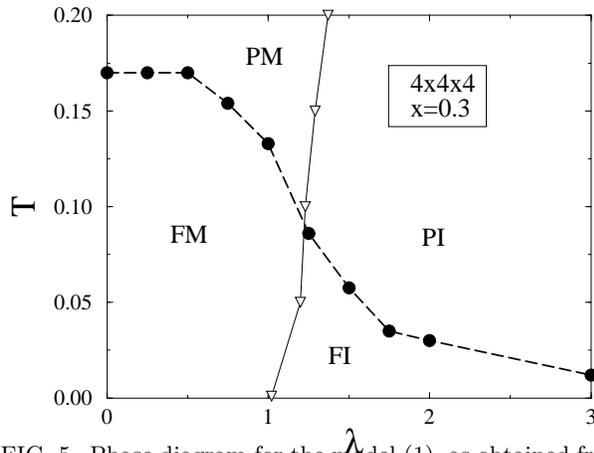, width=\columnwidth, angle=0}
\caption{Phase diagram for the model (\ref{hamil}), as
obtained from the MC simulation on a $4^3$ lattices at $x=0.29$.
Filled dots: the F-P critical $T$. Open-triangles: M-I critical $T$.
}
\label{Fig5}
\end{figure}

The results shown up to now can be understood qualitatively and
semiquantitatively within the few carriers limit.  However, when the
polaron density approaches the percolation threshold of the cubic
lattice ($p_{\mathrm c}=0.31$~\cite{Stauffer}), the polaronic wave
function becomes far less localized. In Fig.~\ref{Fig5}, we show the
phase diagram for $x=0.3$. We find the
same phases as in the $x=0.08$ case.  In the I case, all carriers are
polarons. Notice the absence of a coupling between the P-M and the M-I
transitions, the former being of the second order.  
Therefore in our model, we do not get a $T$-dependent M-I transition
at $x=0.3$. Nevertheless our model only includes an orbital per Mn ion.
The use of a more realistic band structure, will increase the phase
space for the polarons and a $T$-dependent M-I transition will
occur also at higher doping levels.

In summary,
we have studied a DE model coupled with the breathing modes of the
MnO$_6$ octahedra. Given the collective nature of these modes, the
spatial distribution of the lattice distortions is 
inhomogeneous and a Mean Field study difficult. A throughout MC
investigation of the phase diagram has been carried out. Four phases
have been found: FM, FI, PM and PI. A first order M-I
transition with growing $T$ (coincident with the F-M transition) has
been found for the first time on a MC simulation. This transition
survives for a finite range of doping and coupling constant. The M-I
transition is induced by the self-trapping of every carrier on a
polaron, making the Fermi level to lie on the gap between the
polaronic and DE bands (for $t\approx 0.16$eV~\cite{rvw},
the gap will be optical).  We argue that the mechanism presented here
is relevant for the formation of the paramagnetic-insulating phase
experimentally observed in the phase-separated state of colossal
magnetoresistive manganites~\cite{jaime99,heffner00,adams00,wu01}.

We are grateful to F. Guinea, J.L. Alonso and S. Fratini for discussions.
Financial support is acknowledged from grants PB96-0085 (MEC, Spain)
and CAM-07N/0008/2001 (Madrid, Spain). V.M-M. is partially supported by
E.C. contract HPMF-CT-2000-00450.


\begin{references}
\bibitem{Nota} In the case of gap-less systems, more powerful
techniques are available~\cite{Alonso00b}, but for systems with a gap
they have not been tested yet.
\end{references}
\end{document}